\documentclass[aps,prl,twocolumn,showpacs,superscriptaddress]{revtex4}
\input{psfig.sty}

\newcommand{\gtrsim}{ \mathop{}_{\textstyle \sim}^{\textstyle >} }

\newcommand{\vev}[1]{\mbox{$\langle #1 \rangle$}}
\newcommand{\eqn}[1]{&\hspace{-0.5em}#1\hspace{-0.5em}&}

\newcommand{\GeV}{\mbox{GeV}}

\def\sss{\scriptscriptstyle}

\begin{document}

%\preprint{IPT-UNIL-??-2003}

\title{Late Reheating, Hadronic Jets and Baryogenesis}

\author{Takehiko Asaka}
%\email{Takehiko.Asaka@ipt.unil.ch}
\affiliation{Institute of Theoretical Physics, Swiss Federal Institute
of Technology, CH-1015 Lausanne, Switzerland}

\author{Dmitri Grigoriev}
%\email{dima@thphys.may.ie}
\affiliation{Institute for Nuclear Research of RAS, Moscow 117312, Russia}
\affiliation{Mathematical Physics, Natl. Univ. of Ireland Maynooth,
Maynooth, Co. Kildare, Ireland}

\author{Vadim Kuzmin}
%\email{kuzmin@al20.inr.troitsk.ru}
\affiliation{Institute for Nuclear Research of RAS, Moscow 117312, Russia}

\author{Mikhail Shaposhnikov}
%\email{Mikhail.Shaposhnikov@ipt.unil.ch}
\affiliation{Institute of Theoretical Physics, Swiss Federal Institute
of Technology, CH-1015 Lausanne, Switzerland}

\date{October 8, 2003}

\begin{abstract}

If inflaton couples very weakly to ordinary matter the reheating
temperature of the universe can be lower than the electroweak scale.
In this letter we show that the late reheating occurs in a highly
non-uniform way, within narrow areas along the jets produced by
ordinary particles originated from inflaton decays. Depending on
inflaton mass and decay constant, the initial temperature inside the
lumps of the overheated plasma may be large enough to trigger the
unsuppressed sphaleron processes with baryon number non-conservation,
allowing for efficient local electroweak baryogenesis. 

\end{abstract}


\maketitle

%\section{Introduction}
%%%%%%%%%%%%%%%%%%%%%%%%%%%%%%%%%%%%%%%%%%%%%%%%%%%%%%%%%%%%%%

{\em Introduction.}--- The inflationary paradigm (for a review see,
e.g. books \cite{Linde:1990nc})
happened to be very successful for understanding of the basic
properties of the universe. It is assumed that the energy density of
the early universe was dominated by a potential energy of a scalar
field - inflaton. The accelerating expansion of the universe then
leads to a solution of the horizon and flatness problems, whereas
quantum fluctuations of the inflaton field result in  density
perturbations necessary for structure formation and seen as the
temperature fluctuations of the cosmic microwave background
radiation. 

The exponential expansion of the universe during inflation must be
eventually replaced by a radiation dominated epoch, that started  at
least somewhat before nucleosynthesis and lasted till about
recombination. The process of transfer of the energy density of the
inflaton to ordinary matter is usually called reheating, and several
scenarios for how it may proceed have been proposed. Quantitatively,
they can be distinguished by the value of the reheating temperature
$T_R$, below which the universe expansion is dominated by radiation.
If the coupling of inflaton to the ordinary matter is sufficiently
strong, the energy transfer occurs very rapidly due to the phenomenon
of broad parametric resonance right after inflationary stage
\cite{Kofman:1994rk}; this leads to high reheating temperatures $T_R
\sim 10^{10}$ GeV or so. If, on the contrary, the coupling is very
weak, the exponential expansion of the universe is first replaced by
a matter dominated period during which inflaton oscillates without
dissipation. Then, the perturbative decays of inflaton heat the
universe up to some temperature $T_R$. In the latter case the
reheating temperature can be very low, with the only reliable bound
$T_R > 1$ MeV coming from the successful predictions of the big bang
nucleosynthesis. The small values of the reheating temperatures can
naturally occur in certain supergravity models (see, e.g.
\cite{Izawa:1996dv}) where the decay rate of the inflaton is
suppressed by the Planck~scale. 

A successful cosmological model should also explain the absence of
antimatter in the universe and the baryon to entropy ratio $n_B/s
\simeq 9 \cdot 10^{-11}$ \cite{Spergel:2003cb}. 
The baryogenesis must occur after
inflation since otherwise all created baryonic excess will be
exponentially diluted. 

Quite a number of different baryogenesis mechanisms exist in theories with high
reheating temperatures. The explosive particle production during
inflaton decay in the wide resonance case \cite{Kolb:1996jt}  may
lead to production of GUT leptoquarks, the subsequent CP-violating
and baryon number non-conserving decay of which gives rise to baryon
asymmetry of the universe.  The gravitational production of
superheavy particles has the similar effect
\cite{Kolb:1998he}. The thermal
\cite{Buchmuller:2003gz} or non-thermal \cite{Lazarides:wu}
production of heavy Majorana neutrinos may prepare suitable initial
conditions for generating the lepton asymmetry
\cite{Fukugita:1986hr}, which is transformed then to baryon asymmetry
due to anomalous electroweak number non-conservation
\cite{Kuzmin:1985mm}. If the reheating temperature is greater than
the electroweak scale an electroweak baryogenesis (for reviews see
\cite{Rubakov:1996vz}) can take place.

The theories with high $T_R$ may, however, be in conflict with
observations because of overproduction of dangerous relics like
gravitinos \cite{Khlopov:pf}. From this
point of view the theories with small reheating temperature are more
advantageous, as the production of unwanted particles is automatically
suppressed. At the same time, the problem of baryon asymmetry of the
universe in much more difficult if $T_R$ is relatively small, simply
because in this case the low energy baryon number nonconservation is
required. Thus, none of the mechanisms related to the production of
heavy leptoquarks or Majorana neutrinos is operative, and one has to
rely on some variant of electroweak
\cite{Garcia-Bellido:1999sv,Davidson:2000dw}
or Affleck-Dine
\cite{Affleck:1984fy} baryogenesis.

The electroweak baryogenesis, occurring in expanding and almost
equilibrium plasma is highly constrained, as it requires the freezing
of the sphaleron processes after the first order phase transition.
This condition can be converted in the upper bound on the Higgs mass
in the minimal standard model
\cite{Shaposhnikov:jp}. This bound cannot be
satisfied with the experimental value of the top quark mass
\cite{Kajantie:1996mn}, so that new physics is required. In the
minimal supersymmetric standard model there exist a (small)
region of the parameter-space leading to a sufficiently strong first
order phase transition (see \cite{Laine:1998qk} and references therein). 

In \cite{Davidson:2000dw} it was pointed out that the maximum
temperature during reheating can in fact be much larger than the
temperature $T_R$ and that the rate of the universe expansion in the
region of the electroweak phase transition can be faster than it is
usually assumed. This allowed the authors to relax  somewhat the
Higgs mass bound for the electroweak baryogenesis.

In this letter we will show that the late inflaton decays heat up the
plasma in a very non-uniform way and that the {\em local} temperature
along the trajectories of decay products is substantially higher than
the average one.  For a range of parameters the mean temperature of
the plasma is {\em small} enough to shut off the sphaleron transitions
whereas the temperature of the overheated regions is {\em large}
enough to switch them on. Electroweak baryogenesis is then possible in
these regions. The overheated regions cool down due to diffusion and
expansion. This process has a highly non-equilibrium character, so no
first order phase transition is required to satisfy the wash out
condition and thus no bound on the Higgs mass is implied at all.
Remarkably, the resulting baryon asymmetry does not depend much on
many details of the process and may be consistent with the observed
one for large enough CP-violation.
%%%%%%%%%%%%%%%%%%%%%%%%%%%%%%%%%%%%%%%%%%%%%%%%%%%%%%%%%%%%%%%%%%

{\em Local overheating.}---Let $M_\phi$ is the inflaton mass and
$\Gamma_\phi = f_\phi M_\phi$ is its width. Assuming the instantaneous
decay of the inflaton at time $t_\phi=1/\Gamma_\phi$ the reheating
temperature is given by $T_R=\sqrt{f_\phi M_\phi M_0}$, where $M_0 =
M_{Pl}/1.66 g_\ast^{\frac{1}{2}}$, $M_{Pl}$ is the Planck mass,
$g_\ast\sim 100$ is the effective number of massless degrees of
freedom. The typical numbers we will be interested in are $T_R < T_W
\sim 100$ GeV (here $T_W$ is the freezing temperature of the sphaleron
processes), and $M_\phi> 10^{10}$ GeV, what requires rather weak decay
constant $f_\phi \sim 10^{-24}$, or smaller.  We shall assume, for
simplicity, that inflaton decays into quark-antiquark pair, though
other decay channels lead essentially to the same result.

The number density of inflatons decreases with time as $n_\phi a^3 = n_0
\exp(-\Gamma_\phi t)$ with scale factor $a$. 
The first decays occur essentially in the
vacuum, whereas at $t\sim t_\phi$ inflatons are surrounded by the
plasma with the temperature $T \sim T_R$. The decay products of
inflaton, ultra-relativistic quarks with the energy $M_\phi/2 \gg
T_R$, are injected into the plasma and heat it locally. Our aim now
is to understand the typical size and geometry of the overheated
regions, as well as their temperature. 

The dynamics of energy losses of high energy quarks and gluons in
quark-gluon plasma is a complicated problem which must incorporate
the Landau-Pomeranchuk-Migdal effect
\cite{Landau:um} and non-abelian
character of interaction of quarks and gluons. The main features of
it have been understood only recently 
(see \cite{Baier:1994bd} and references therein), the crucial one being 
that the energy loss per unit length is increasing in infinite 
plasma as a square root of parton energy, see eq.~(\ref{Ltot}) below.

The most significant part of energy losses is related to the soft
gluon emission in the multiple scattering of the hard parton on the
particles of the plasma. The energy spectrum $I$ of the emitted
gluons per unit length has the following approximate form for
$\omega_{\rm \sss BH} \ll \omega \le E_0$ : 
\begin{eqnarray}
  \label{SPE}
  \omega \frac{d^2 I}{ d\omega dz }  =
  \frac{2}{\pi} 
  \frac{ \alpha_s }{ \lambda_g }
   \sqrt{ \displaystyle
        \frac{\omega_{\rm \sss BH}}{\omega}
        \ln \left( \frac{\omega}{\omega_{\rm \sss BH}} \right)}~,
\end{eqnarray}
where $\omega$ and $E_0$ denote the gluon and parton energies, 
$\alpha_s$ is QCD coupling constant, $\lambda_g$ is the mean free
path of the gluon, and $\omega_{\rm \sss BH}$ is the Bethe-Heitler
frequency, $\omega_{\rm \sss BH} = \lambda_g \mu^2$. Here $\mu$ is
the typical screening mass which is assumed to be of the order of the
Debye mass in the plasma. From (\ref{SPE}) one gets the stopping
distance of the initial parton $L_{\rm \sss tot}$ :
\begin{eqnarray}
  \label{Ltot}
  L_{\rm \sss tot} 
  = \frac{\pi}{2} \frac{ \lambda_g }{ \alpha_s }
  \sqrt{ \frac{E_0}{\widetilde \omega_{\rm \sss BH}} }~,
\end{eqnarray}
where $\widetilde \omega_{\rm \sss BH} \simeq \omega_{\rm \sss BH}
\log \left( {E_0}/{ \omega_{\rm \sss BH} } \right)$ for
$E_0\gg\omega_{\rm \sss BH}$.
The average energy of the emitted gluons is 
$\vev{\omega} = \sqrt{ \omega_{\rm \sss BH} E_0 }/{ 2 }$
and their number is given by 
$N_g = 2 \sqrt{ { E_0 }/{ \omega_{\rm \sss BH} } }$.

If the emitted gluons have energies  $\omega \gg \omega_{\rm \sss
BH}$, they lose their energies  as the parent parton does.  This
cascade process will terminate  when the energy of the emitted gluons
becomes comparable to or smaller than $\omega_{\rm \sss BH}$.
We find the energy for the $n$-th gluons as
\begin{eqnarray}
  \vev{\omega}_n = 
  \frac{1}{4^{1-1/2^n}} \omega_{\rm \sss BH}   
  \left( \frac{E_0}{\omega_{\rm \sss BH}} \right)^{1/2^n}~,
\end{eqnarray}
so number $N_{BH}$ of the cascade steps from $\vev{\omega}_0 = E_0$ to
$\vev{\omega}_{N_{\rm \sss BH}} \le \omega_{\rm \sss BH}$ is typically
$3-4$ for $E_0 \sim 10^{10}-10^{11}$ GeV.     

To find the geometry and volume of the region where the emitted
gluons deposit their energy we note that the radiation with the
frequency $\omega$ is mainly concentrated in the cone with a small
angle $\theta$, which can be computed with the help of the results of
\cite{Baier:2001qw} and is given by
\begin{eqnarray}
  \theta_\omega{}^2 
   \simeq 
  \frac{ L_\omega ~ \overline \omega_{\rm \sss BH } } 
  { \lambda_g^2 ~ \omega^2 }~,
\end{eqnarray}
where $L_\omega$ is length of the energy loss for the gluon with energy
$\omega$, 
\begin{eqnarray}
  L_\omega =
  \frac{2 \pi}{9} \frac{ \lambda_g }{ \alpha_s }
  \sqrt{ \frac{\omega}{\overline \omega_{\rm \sss BH} } } ~,
\end{eqnarray}
$\overline \omega_{\rm \sss BH} = \omega_{\rm \sss BH} \ln \left(
{\omega}/{\omega_{\rm \sss BH}} \right)$. 
The transverse distance~$r_\omega = L_\omega\theta_\omega$ 
traveled by a gluon with the energy $\omega$ is then 
\begin{eqnarray}
  r_\omega \simeq
  \left( \frac{2\pi}{9 \alpha_s} \right)^{3/2}
  \frac{  \lambda_g^{1/2}  }{ \omega_{\rm \sss BH}^{1/4} ~\omega^{1/4} } 
  \times
   \left[  \ln \left( \frac{ \omega }{ \omega_{\rm \sss BH } } 
      \right) \right]^{-1/4}~,
\end{eqnarray}
(the log factor should be neglected at $\omega\sim\omega_{\rm \sss BH}$).
It is seen that, although the cascade process produces gluons having
various energies, the largest $r_\omega$ is due to the gluons with the
lowest energy  $\omega \simeq \omega_{\rm \sss BH}$. Thus, the
overheated region has an approximate cylindrical form with the length
given by eq. (\ref{Ltot}) and with the radius $r_\ast 
\simeq r_{\omega_{\rm \sss BH}}$. Its volume is $V = \pi
r_\ast^2 L_{\rm \sss tot}$.

Now we are at the point to estimate an effective temperature inside
the cylinder in which the emitted gluons are living. By assuming the
energy conservation and the rapid thermalization, we obtain that the
effective temperature $T_\ast$, found from 
$\frac{\pi^2 g_\ast}{30} T_\ast^4 = \frac{E_0}{V}$,
is given by
\begin{eqnarray}
\label{Teff1}
  T_\ast \simeq 5.3 \times 10^{-2}
  \left( \frac{100}{g_\ast} \right)^{1/4}
  \mu^{3/4} \left[ \frac{E_0}{\lambda_g} 
  \ln \left( \frac{ E_0 }{ \omega_{\rm \sss BH} } \right) \right]^{1/8}.
\end{eqnarray}
For numerical estimates we take the non-perturbative value of the
Debye screening mass found in \cite{Kajantie:1997pd}. As for the gluon
mean free path, we take the value found in \cite{Wang:1994fx} and
denoted as $\lambda_g^{\rm \sss el}$ or simply take a gluon damping
rate $\gamma_g$ from \cite{Braaten:1990it}. They are different by
roughly one order of magnitude, what allows one to get an estimate of the
uncertainties in (\ref{Teff1}).

\begin{figure}
\centerline{\psfig{figure=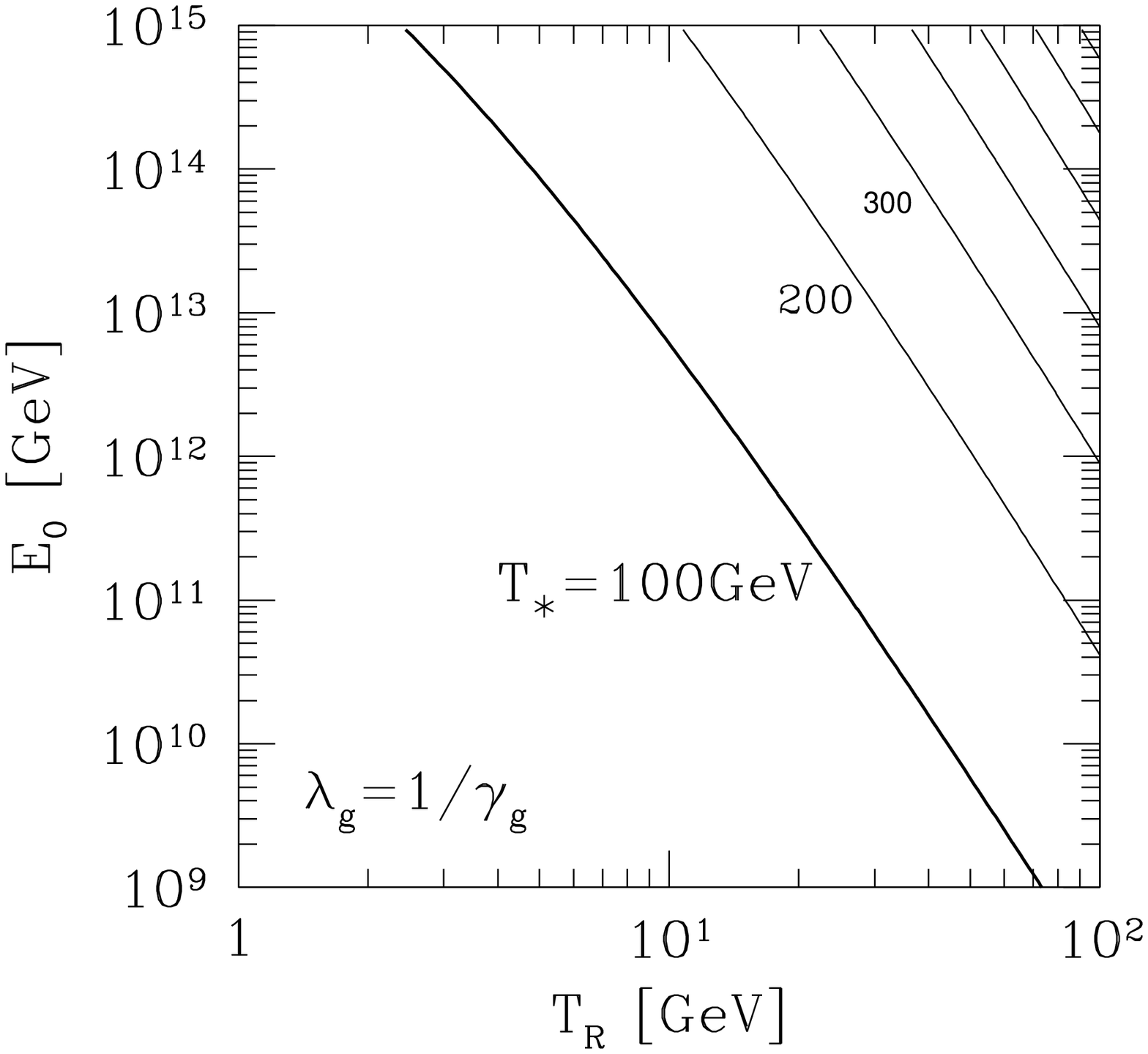,height=7.5cm}}
\vspace{-1cm}
\centerline{\psfig{figure=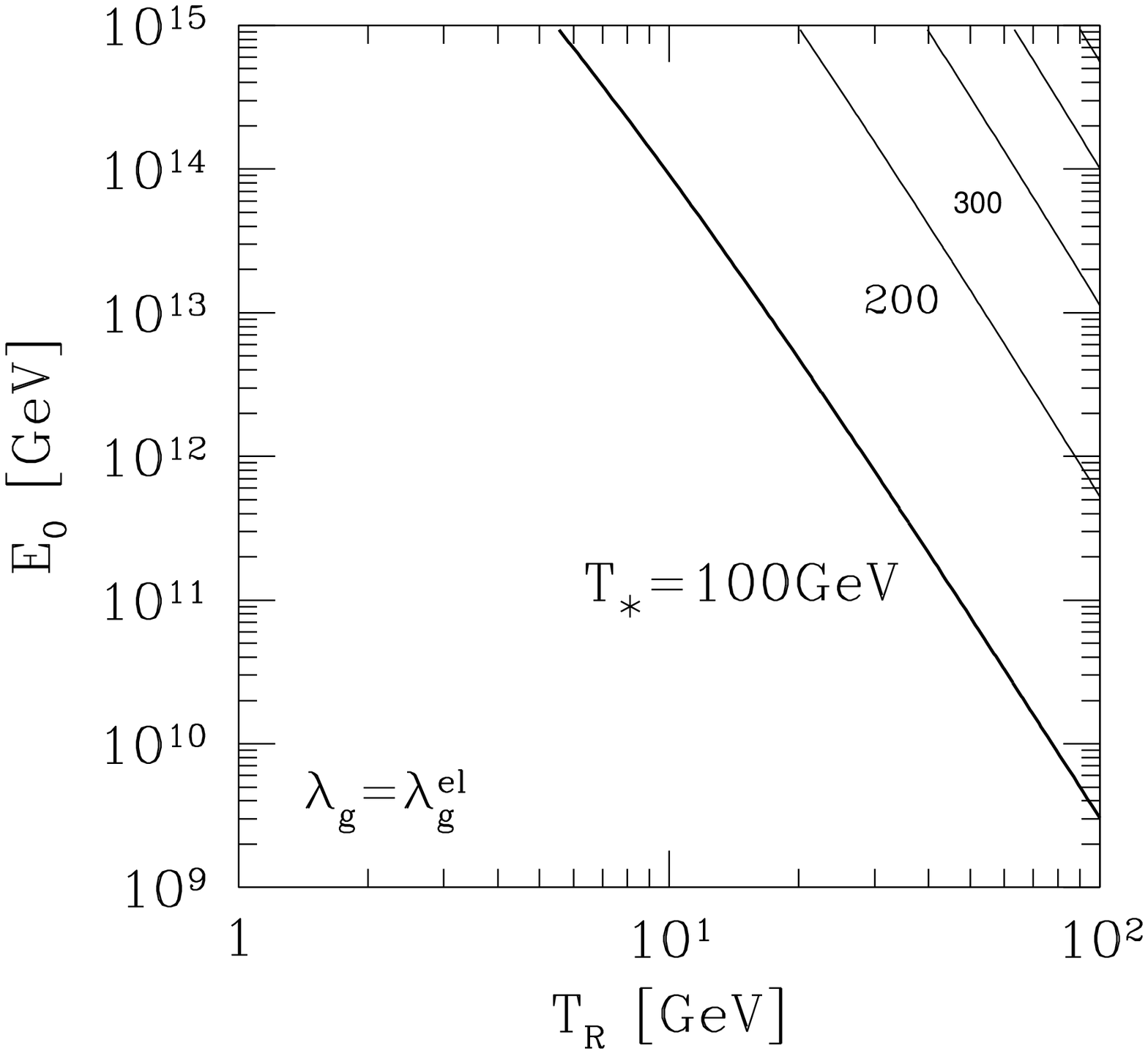,height=7.5cm}}
\vspace{-0.5cm}
  \caption{Contour plots of effective temperature $T_\ast$
    in the $T_R$-$E_0$ plane for $\lambda_g = 1/\gamma_g$ (up) and
    $\lambda_g = \lambda_g^{\rm \sss el}$ (down).}
  \label{fig:Teff}
\end{figure}

For example, when $T_R = M_Z$ and $E_0=10^{11}$ GeV, we find (cf.
Fig.~\ref{fig:Teff}) that the effective temperature is $T_\ast \simeq
212~\GeV$  ($\lambda_g = 1/\gamma_g$) or $T_\ast \simeq 151~\GeV$ 
($\lambda_g = \lambda_g^{\rm \sss el}$). 
Fig.~\ref{fig:Teff} clearly demonstrates 
that the local
temperature of the overheated regions can substantially exceed the
freezing temperature of the sphaleron processes, provided the
background cosmic temperature is relatively high, say $T_R \gtrsim   
10~\GeV$ and the energy of the parent parton is extremely high, say
$E_0 \gtrsim 10^{10}$ GeV.

%%%%%%%%%%%%%%%%%%%%%%%%%%%%%%%%%%%%%%%%%%%%%%%%%%%%%%%%%%%%%%%%%%%%
{\em Baryogenesis.}---Let us assume now that the parameters of the
inflaton are such that the temperature $T_\ast$ is high enough: 
$T_\ast > T_{\rm \sss sph} \sim 100~\GeV > T_W$.  
Here $T_{\rm
  \sss sph}$ is the temperature above which the rate of the sphaleron
transitions is unsuppressed and is given by (per unite time and unite
volume) $\Gamma_{\rm \sss sph} \simeq \kappa \alpha_W^5 T_\ast^4$,
where $\kappa \sim 10$ \cite{Moore:2000mx}.  At the same time, the
reheating temperature $T_R$ can be small enough, so that baryon number
is conserved away from the overheated regions ({\it i.e.} $T_R <
T_W$).  This highly non-equilibrium situation is possible at any
choice of parameters of the underlying electroweak theory, and,
therefore, the wash out bound of
\cite{Shaposhnikov:jp} is not applicable.

The baryon asymmetry of the universe can be estimated as the
number of sphaleron transitions which takes place 
inside the overheated regions and go asymmetrically due to 
CP-violation:
\begin{eqnarray}
  \frac{n_B}{s} \eqn{\simeq}
  \frac{n_{\rm parton}}{s}
  \times \kappa \alpha_W^5 T_\ast^4 ~ V~ \Delta t 
  \times \delta_{\rm \sss CP}~,
\end{eqnarray}
where $n_{\rm parton}$ is the number density of the highly energetic
partons coming from the inflaton decay, $\Delta t$ denotes the
lasting time of the rapid shaleron processes and we have introduced
$\delta_{\rm \sss CP}$ to represent the effective magnitude of
CP-violation. 

With the two-particle decays of the inflaton the number of partons is
simply $n_{\rm \sss parton} = 2 n_\phi$, and  $\frac{n_\phi}{s}
\simeq \frac{3}{4} \frac{T_R}{M_\phi}$ from the condition defining
the reheating temperature. Putting all factors together, we get
\begin{eqnarray}
  \frac{n_B}{s} 
 \simeq 10^{-8} ~T_R~ \Delta t ~\delta_{\rm \sss CP}~.
 \label{bau}
\end{eqnarray}
Quite amazingly, besides the expected CP-violating factor, the
result depends just on the reheating temperature and on the sphaleron
transition time. In particular, the temperature of the
overheated regions has canceled out from eq. (\ref{bau}). What is
important is that the overheated regions must be in the symmetric
phase of the electroweak theory, where baryon number non-conservation
is not suppressed.

The overheated regions cool down by the growth of the volume due to
diffusion, and the rapid spharelon processes terminate eventually when
the temperature inside becomes $T_{\rm \sss sph}$.  From the energy
conservation the radius of the overheated region at $T_{\rm \sss sph}$ 
is given by $\ell = r_\ast (T_\ast/T_{\rm \sss
  sph})^2$ and is reached after the diffusion time $ \Delta t\simeq \ell^2 /4D$
(we find from \cite{Arnold:2000dr} that the diffusion coefficient
$D \sim 1/\gamma_g \sim 0.1 \lambda_g^{\rm \sss el}$),
which gives 
$\Delta t \simeq \frac{ r_\ast^2 }{ 4 D }  
\left( \frac{T_\ast}{T_{\rm \sss sph}} \right)^4$.
For a region of parameters $\Delta t$ is long enough 
to thermalize $W$-bosons
which are essential to the sphaleron processes.   

Finally, one gets 
\begin{eqnarray}
\frac{n_B}{s} \gtrsim (1-10)\cdot 10^{-7}
\delta_{\rm \sss CP},
\end{eqnarray}   
depending on the estimate of the gluon mean-free path discussed
above. So, the extension of the standard model with suitable
CP-violation may work.

{\em Conclusions.}---We have shown that a successful electroweak
baryogenesis can take place in inflation models with low reheating
temperature. Though our estimates are rather rude (even parton
losses in plasma have rather large uncertainties due to the lack
of exact knowledge of the kinetic coefficients), it is clear
that the increase of the inflaton mass makes the temperature of the
overheated regions higher, what triggers the mechanism at some
critical inflaton mass.

Generally speaking, the plasma overheating by the processes discussed above
will take place along the trajectories of decay products of any sufficiently
heavy particles provided their decay products interact 
with 
the background plasma.
%%%%%%%%%%%%%%%%%%%%%%%%%%%%%%%%%%%%%%%%%%%%%%%%%%%%%%%%
\begin{acknowledgments}
  We thank D. Kharzeev and U. Wiedemann for very useful discussions on
  the parton energy losses in quark-gluon plasma 
  and P. Tinyakov for
  helpful remarks.  This work was supported by the Swiss Science
  Foundation grant. The work of T.  Asaka was supported by the Tomalla
  foundation. D.G. and V.K. thank Institute of Theoretical Physics of
  Lausanne University, where part of this work was done, for
  hospitality.
\end{acknowledgments}

\end{document}